# Bias and high-order galaxy correlation functions in the APM Galaxy Survey

ENRIQUE GAZTAÑAGA[1] & JOSHUA A. FRIEMAN[2,3]

[1]*Department of Physics, Astrophysics*
*University of Oxford, Nuclear & Astrophysics Laboratory*
*Keble Road, Oxford, England OX1 3RH*

[2]*NASA/Fermilab Astrophysics Center*
*Fermi National Accelerator Laboratory*
*Batavia, IL 60510-0500, USA*

[3]*Department of Astronomy and Astrophysics*
*University of Chicago, Chicago, IL 60637*

## ABSTRACT

On large scales, the higher order moments of the mass distribution, $S_J = \overline{\xi}_J / \overline{\xi}_2^{J-1}$, e.g., the skewness $S_3$ and kurtosis $S_4$, can be predicted using non-linear perturbation theory. Comparison of these predictions with moments of the observed galaxy distribution probes the bias between galaxies and mass. Applying this method to models with initially Gaussian fluctuations and power spectra $P(k)$ similar to that of galaxies in the APM survey, we find that the predicted higher order moments $S_J(R)$ are in good agreement with those directly inferred from the APM survey *in the absence of bias*. We use this result to place limits on the linear and non-linear bias parameters. Models in which the extra power observed on large scales (with respect to standard CDM) is produced by scale-dependent bias match the APM higher order amplitudes only if non-linear bias (rather than non-linear gravity) generates the observed higher order moments. When normalized to COBE DMR, these models are significantly ruled out by the $S_3$ observations. The cold plus hot dark matter model normalized to COBE can reproduce the APM higher order correlations if one introduces non-linear bias terms, while the low-density CDM model with a cosmological constant does not require any bias to fit the large-scale amplitudes.

*Subject Headings*: Large-scale structure of the universe — galaxies: clustering

# 1 Introduction

In standard models for galaxy formation, such as cold dark matter (CDM) and its offshoots, it is usually assumed that the observable galaxy distribution is related through a simple bias mechanism to the underlying matter distribution predicted by theory (e.g., Bardeen, *et al.* 1986). In essence, following Kaiser (1984a,b) and Bardeen (1984), one assumes that galaxies form from peaks above some global threshold in the smoothed linear density field. In the limit of high threshold and small variance, the peaks model is well approximated by the commonly employed linear bias scheme, in which the galaxy and mass density fields, $\delta_g(\boldsymbol{x}) = (n_g(\boldsymbol{x}) - \bar{n}_g)/\bar{n}_g$ and $\delta(\boldsymbol{x}) = (\rho(\boldsymbol{x}) - \bar{\rho})/\bar{\rho}$, are related through a constant bias factor,

$$\delta_g(\boldsymbol{x}) = b_g \delta(\boldsymbol{x}) \quad . \tag{1}$$

More generally, if the bias is *local*, the galaxy field may be some non-linear function of the density field, $\delta_g(\boldsymbol{x}) = f(\delta(\boldsymbol{x}))$. Recently, more complex, *non-local* schemes for biasing have also been studied (Babul and White 1991, Bower, *et al.* 1993); in these models, the effective bias factor becomes scale-dependent. If the bias factor increases with scale, the galaxy spectrum will have more relative power at large scales. Indeed, it has been shown that such modifications of the standard bias scheme can generate the excess large-scale power found, e.g., in the APM survey (Maddox *et al.* 1990) within the context of the standard ($\Omega h = 0.5$) CDM model. This modification of the standard CDM scenario is fundamentally different from other CDM variants, such as non-zero $\Lambda$, tilt, or cold plus hot dark matter, where there is genuine extra power in the density field.

To date, two-point statistics (correlation functions and power spectra) have been the dominant benchmark for testing theories of biased structure formation. It is difficult, however, to test biasing models with two-point statistics alone, because one must specify a number of input variables to predict the amplitude of the power spectrum, e.g., the density of the universe $\Omega$, the Hubble constant $h$ (where $H_0 = 100h$ km s$^{-1}$ Mpc$^{-1}$), the type of dark matter, the cosmological constant $\Lambda$, the initial amplitude and the clustering interactions (gravity). On the other hand, the non-linear perturbation theory predictions for the higher order moments of the mass distribution, $S_J = \bar{\xi}_J/\bar{\xi}_2^{J-1}$, are approximately independent of time, scale, density, or geometry of the cosmological model (Juskiewicz, Bouchet, and Colombi 1993, Bernardeau 1994). The higher moments only depend on the hypothesis that the initial fluctuations are small and quasi-Gaussian and that they grow via gravitational clustering. Thus, the moments $S_J$ are an excellent tool for discriminating among biasing scenarios.

Frieman & Gaztañaga (1994) have considered how the higher order irreducible moments of the galaxy distribution can be used as a test of models for large-scale structure. In particular, they studied the three-point function in the standard CDM model and its variants with extra large-scale power (e.g., $\Omega h = 0.2$ CDM), as well as in a generalized version of the non-local, scale-dependent bias scheme embodied in the cooperative galaxy formation (hereafter CGF) model of Bower *et al.* . Using the results of second-order perturbation theory (Fry 1984), they compared the predictions of these models for the three-point function $\xi_3$ with data from the Center for Astrophysics (Huchra, *et al.* 1983), Southern Sky (Da Costa, *et al.* 1991), and Perseus-Pisces (Haynes and Giovanelli 1988) redshift surveys in the mildly non-linear regime ($\xi_2 < 1$).

In this *Letter*, we compare these model predictions to the higher order galaxy correlations recently estimated for the APM survey by Gaztañaga (1994). The values for the $S_J$ moments in the APM survey are statistically more reliable than those inferred from the redshift surveys above, because the APM catalog covers over 200 times more volume and contains 300 times more galaxies than the combined CfA/SSRS. The skewness $S_3$ found in the APM survey is typically larger than in the optical redshift surveys (Gaztañaga 1992). However, the disagreement is only significant on small scales: at the largest scales probed by the redshift surveys, $R \simeq 20\,h^{-1}\,\mathrm{Mpc}$, the value of $S_3 \sim 2$ found in the CfA/SSRS catalogs is in good agreement with the APM results. Furthermore, comparison of the CfA and SSRS catalogs with comparable sub-samples of the APM indicates that the difference in skewness on small scales is attributable to the fact that the local galaxy distribution traced by the CfA and SSRS catalogs does not correspond to a "fair sample" rather than to uncertainties in the APM selection function (Gaztañaga 1994).

## 2 The APM galaxy moments

In the weakly non-linear regime ($\overline{\xi}_2 < 1$), non-linear perturbation theory has been shown to hold remarkably well when compared with N-body simulations (Baugh, Gaztañaga, & Efstathiou in preparation, Efstathiou *et al.* 1988, Bouchet and Hernquist 1992, Weinberg & Cole 1992, Juskiewicz, Bouchet, & Colombi 1993, Bernardeau 1994). In perturbation theory (Fry 1984, Goroff, *et al.* 1986, Bernardeau 1992, 1994), the matter density field evolved gravitationally from Gaussian initial conditions leads to a hierarchical clustering pattern of the form $\overline{\xi}_J = S_J\,\overline{\xi}_2^{J-1}$. Here $\overline{\xi}_J$ is the volume-averaged $J-$point correlation function,

$$\overline{\xi}_J(V) = \frac{1}{V^J} \int \cdots \int d^3r_1 ... d^3r_J \xi_J(r_1,...,r_J) W(\boldsymbol{r}_1)...W(\boldsymbol{r}_J) \quad , \qquad (2)$$

where $W(\boldsymbol{r})$ is a window function of characteristic volume $V$ (below, we use a top-hat window of radius $R$). For models without strong features in the linear power spectrum (e.g., CDM and its popular variants), the perturbative $J-$point moments $S_J$ depend only weakly on the window smoothing scale (for power law spectra, the $S_J(R)$ are independent of $R$). Here we focus on the hierarchical amplitudes $S_3$ and $S_4$, the normalized measures of the skewness and kurtosis of the smoothed density probability distribution function.

Figure 1 shows the values of $S_3(R)$ and $S_4(R)$ for spherical cells of radius $R$, estimated from the angular amplitudes $s_3(\theta)$ and $s_4(\theta)$ in the APM galaxy survey (cf. Fig. 4 of Gaztañaga 1994). The error-bars correspond to the one-sigma dispersion between four disjoint zones of the APM map and are therefore an estimate of the 'cosmic variance'. The bins in the data correspond to the bins in the angular data: the original angular cell counts are done in conical volumes. Each angular bin in the $s_J(\theta)$ data is such that the conical volume in one cell is at least 100% larger than the volume of the next smallest bin, so the correlations in each bin are roughly independent from the others. The angular amplitudes have been corrected by a factor $0.95^{J-2}$ to account for residual contamination from merged images in the APM survey (Maddox, *et al.* 1990, Gaztañaga 1994); since this factor is approximate, the error-bars on $s_J(\theta)$ were constrained to be at least as large as the correction factor. We have also performed N-body simulations to verify that the errors introduced by uncertainties in the APM selection function and in the two- to three-dimensional inversion factors are small compared with those above (Gaztañaga 1994, and in preparation). Thus, for the purposes of comparison with theoretical models, the data points shown in Fig. 1 can be taken to be independent and have been assigned conservative errors. (For additional details, see Gaztañaga 1994.)

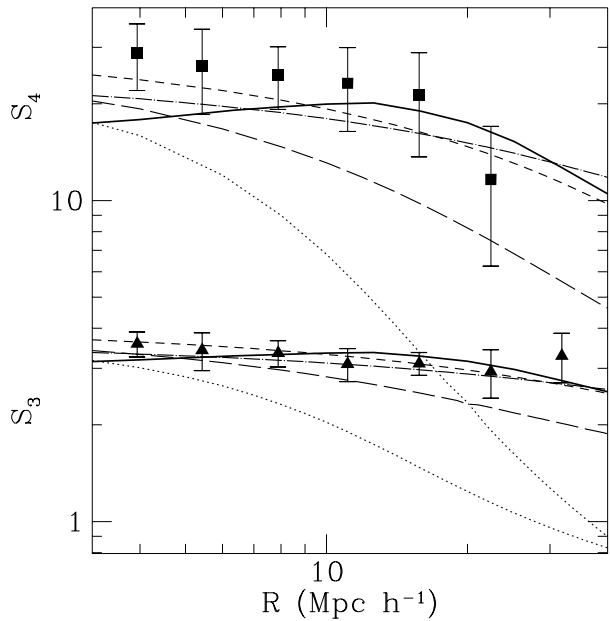

Figure 1: The volume-averaged skewness $S_3(R)$ and kurtosis $S_4(R)$ for spherical cells of radius $R = \theta \mathcal{D}$, estimated from $s_J(\theta)$ in the APM survey, where the effective depth $\mathcal{D} \simeq 400\,h^{-1}\,\mathrm{Mpc}$. Error-bars correspond to one-sigma dispersion of $s_J(\theta)$ between 4 zones of the APM map. Curves correspond to the $S_J(R)$ predictions in perturbation theory for standard CDM ($\Gamma = 0.5$, long-dash curves), low-density CDM ($\Gamma = 0.2$, short-dash curves), the inferred APM spectrum (solid curves) and standard CDM modified by the scale-dependent bias (CGF model), with $\kappa = 2.29$, $R_s = 20\,h^{-1}\,\mathrm{Mpc}$ (dotted curve). These models are all shown with $b_1 \equiv b = 1$, $c_2 = c_3 = 0$. Also shown (dot-dash curves) are the moments for a biased $\Gamma = 0.2$ model with $b = 1.5$, $c_2 = 0.45$, $c_3 = 0.2$ choosen to give similar predictions to the unbiased case.

For the models, we assume the initial fluctuations are small and Gaussian, and that they evolve under the influence of gravity in an expanding universe. We estimate the moments $S_J(R) = \overline{\xi}_J(R)/\overline{\xi}_2(R)^{J-1}$ for the matter distribution using $(J-1)$-order perturbation theory (Fry 1984, Bernardeau 1994). The results for $S_J$ are completely determined by the linear power spectrum $P(k)$, for which we adopt the parametric form

$$P(k) = A\sigma_8^2 k \left(1 + \frac{1.7k}{\Gamma} + \frac{9k^{3/2}}{\Gamma^{3/2}} + \frac{k^2}{\Gamma^2}\right)^{-2}, \qquad (3)$$

where the wavenumber $k$ is in units of $h\,\mathrm{Mpc}^{-1}$, and $\sigma_8^2 = \overline{\xi}_2(R = 8\,h^{-1}\,\mathrm{Mpc})$ is the variance of the linear mass fluctuation within (top-hat) spheres of radius $8\,h^{-1}\,\mathrm{Mpc}$. For CDM models, equation (3) gives an accurate fit to the linear power spectrum with $\Gamma = \Omega h$ (Davis, etal. 1985). We consider two models: 'standard CDM' with $\Gamma = 0.5$, and 'low-density CDM' with $\Gamma = 0.2$. For each model, we use the power spectrum $P(k)$ to calculate the variance $\overline{\xi}_2(R)$ and its derivatives $\gamma_n = d^n \log \overline{\xi}_2(R)/d \log^n R$ and use the results of Bernardeau (1994) to calculate the predicted amplitudes $S_3(R)$ and $S_4(R)$. These are shown as the long and short-dash curves in Fig. 1 for standard and low-density CDM. For comparison, the solid curves in Fig. 1 show the predicted higher moments using the power spectrum inferred directly from the APM catalog by Baugh & Efstathiou (1993) as the linear input for higher order perturbation theory. This spectrum is reasonably well approximated by the $\Gamma = 0.2$ CDM model.

To show the effects of scale-dependent bias, we also consider the non-local bias model of Bower, etal. (1993). In this model, the linear galaxy power spectrum is just the standard $\Gamma = 0.5$ CDM spectrum multiplied by the 'cooperative' bias factor $b^2(k) = (1 + \kappa e^{-(kR_s)^2/2})^2$. In the parameter range studied by Bower, etal., the choice $\kappa = 2.29$, $R_s = 20\, h^{-1}$ Mpc for the strength and range of cooperative effects gives the best fit to the APM angular correlation function, and for illustration we focus on this case. The corresponding skewness and kurtosis are shown as the dotted curves in Fig. 1. As noted by Frieman & Gaztañaga (1994), in this model there is a sharp downturn in the $S_J$ on large scales, $R > 10\, h^{-1}$ Mpc, which is at variance with the relative flatness of the observed moments.

The model curves in Fig. 1 were all plotted assuming the linear bias model of equation (1) with $b = 1$. When considering higher-order perturbations, however, one should self-consistently allow for higher-order (non-linear) bias, and replace (1) with an expansion of the form

$$\delta_g = f(\delta) = \sum_{k=1}^{\infty} \frac{b_k}{k!} \delta^k \quad . \tag{4}$$

Fry and Gaztañaga (1993) have shown that such a local transformation $\delta \to \delta_g$ preserves the hierarchical nature of the matter distribution in the limit of small fluctuations, i.e., on large scales, although the values of the resulting galaxy hierarchical amplitudes $S_J^g$ will in general differ from those of the density field (see also Juszkiewicz, etal. 1993). For $J = 3, 4$, they find

$$S_3^g = b^{-1}[S_3 + 3c_2] + \mathcal{O}\langle \delta^2 \rangle \quad , \tag{5}$$

$$S_4^g = b^{-2}[S_4 + 12c_2 S_3 + 4c_3 + 12c_2^2] + \mathcal{O}\langle \delta^2 \rangle \quad , \tag{6}$$

where $S_J^g$ are the galaxy amplitudes, $S_J$ are the matter density moments, $b = b_1$ is the linear bias, and $c_2 = b_2/b$ and $c_3 = b_3/b$ are the lowest relative non-linear bias terms in equation (4). Since $S_3$ and $c_2$ may generally be of order unity, the contribution of the quadratic bias to $S_3^g$ is comparable to that from the linearly biased second-order skewness of the matter. Thus, it is inconsistent to assume either a purely linear bias or to ignore the gravitationally induced skewness, even in the limit of very small fluctuations. This method has been extended to scale-dependent bias models by Frieman and Gaztañaga (1994) and we make use of their results below.

Detailed analysis of N-body results (Baugh, Gaztañaga & Efstathiou 1994 in preparation) show that the perturbation theory predictions for $S_J$ agree remarkably well with the non-linear field for scales where $\overline{\xi}_2 \leq 1$, while on smaller scales, $S_3$ in the simulations rises above the perturbative results. Thus, in Fig. 1, the comparison with perturbation theory must be restricted to scales larger than $R \simeq 7\, h^{-1}$ Mpc to be in the quasi-linear regime. This restriction also implies that we only need include terms up to order $J - 1$ for $S_J$ in the local bias expansion (4). Over this range of scales, Fig. 1 shows that the APM skewness and kurtosis are reasonably well fit (within the errors) by the low-density CDM model with *no* bias, i.e., $b_1 = 1$, $b_{k>1} = 0$, whereas the unbiased CGF model does not agree with the data.

Table 1: Best fit values of the non-linear bias parameters $c_2$ and $c_3$ for different values of $b$, fitted using the APM $S_3$ and $S_4$ and the predictions of different models. Each entry is a range of allowed values $[A_{max}, A_{min}]_{\chi^2}$ where $\chi^2$ is the goodness of the fit at the ends of the range.

| $\Gamma = 0.2$ | $b = 1$ | $b = 1.5$ | $b = 2$ |
|---|---|---|---|
| $c_2$ | $[-0.05, 0.02]_{2.5}$ | $[0.42, 0.49]_{2.5}$ | $[0.57, 0.65]_{2.5}$ |
| $c_3$ | $[-0.3, 2.0]_6$ | $[-0.2, 0.6]_6$ | $[-0.0, 0.8]_6$ |
| $\Gamma = 0.5$ | $b = 1$ | $b = 1.5$ | $b = 2$ |
| $c_2$ | $[0.13, 0.20]_4$ | $[0.29, 0.38]_{2.5}$ | $[0.47, 0.56]_{2.5}$ |
| $c_3$ | $[-0.2, 1.8]_6$ | $[0.5, 1.8]_6$ | $[0.6, 1.9]_6$ |
| CGF | $b = 1$ | $b = 1.5$ | $b = 2$ |
| $c_2$ | $[0.44, 0.48]_{11}$ | $[0.65, 0.70]_4$ | $[0.73, 0.78]_{2.5}$ |
| $c_3$ | $[-0.3, 1.0]_{14}$ | $[0.6, 1.6]_6$ | $[-0.0, 2.1]_6$ |

We can use the five APM amplitudes $S_3(R)$ at $R > 7\,h^{-1}\,{\rm Mpc}$ to find the best-fit quadratic bias parameter $c_2$ for each model shown in Fig. 1, given a value of the linear bias $b$. Table 1 shows the range of values of $c_2$ which give the best fit for each model. The goodness of the fit, parametrized by the $\chi^2(4)$ value, is shown as a subscript to the $c_2$ interval, corresponding to $4 = 5 - 1$ degrees of freedom. We then use a combined $\chi^2$ test with the $S_3(R)$ and $S_4(R)$ APM values to constrain $c_3$, given the range allowed for $c_2$ in the $S_3(R)$ test for each model. In this case we have 9 data points at $R > 7\,h^{-1}\,{\rm Mpc}$ (5 from $S_3$ and 4 from $S_4$) and $7 = 9 - 2$ degrees of freedom (since $c_3$ is free and $c_2$ is allowed to vary within a range). The best-fit ranges for $c_3$ are also shown in Table 1 and correspond to a goodness of fit $\chi^2(7) \simeq 6$, except for the CGF model with $b = 1$ which has $\chi^2(7) \simeq 14$.

For no linear bias, $b = 1$, the $\Gamma = 0.2$ model gives a good fit to the higher moments (at the 1-sigma level) with no non-linear bias, that is, for $c_2 = 0$ and $c_3 = 0$, while the $\Gamma = 0.5$ standard CDM model requires $c_2 \simeq 0.15$ and $c_3 \simeq 1$, and even then only fits the data at the 2-sigma level. The CGF model does not fit the data with $b = 1$ for any choice of the $c_k$ ($\chi^2(4) > 11$). For $b > 1$, on the other hand, all three models can fit the data with different combinations of non-linear bias parameters $c_2$ and $c_3$ as shown in Table 1. However, fits to the data with large values of $b$ are in a sense *ad hoc*, because the agreement is obtained by suppressing the gravitational contribution to $S_J$ and then fitting with the non-linear bias terms alone. These values do not really reflect agreement between the data and the model, but rather the possibility that, in any model, the observed, nearly constant moments can be reproduced by the constant non-linear bias terms in (5) and (6).

In Fig. 2, we show the values of $S_J^g$ obtained directly from the higher order correlations in the APM survey for $J = 3 - 6$. Again, there is reasonable agreement with the unbiased perturbation theory predictions for the $\Gamma = 0.2$ model and for the APM input spectrum from Baugh and Efstathiou (1993)

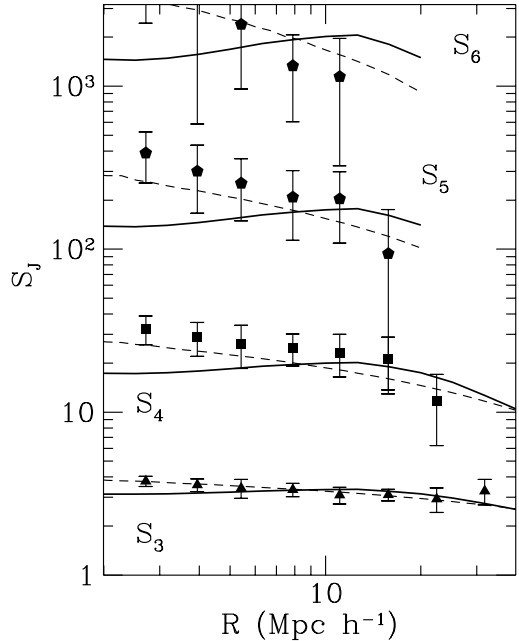

Figure 2: Hierarchical amplitudes $S_J$ for $J = 3 - 6$ compared with the predictions in non-linear perturbation theory with no biasing. Curves correspond to the models in Fig. 1.

on scales $R > 7\,h^{-1}\,\mathrm{Mpc}$.

## 3 Conclusion

The simplest interpretation of the higher order clustering data, in particular the agreement between the matter amplitudes $S_J$ predicted in non-linear perturbation theory and the observed galaxy amplitudes $S_J^g$ in the APM Survey (Figs. 1 and 2), is that there is no bias on scales $R > 7\,h^{-1}\,\mathrm{Mpc}$: the parameters in expansion (4) of $\delta_g = f(\delta)$ are just $b = b_1 \simeq 1$ and $b_{k>1} \simeq 0$. Otherwise, one either has to invoke a delicate, model-dependent balancing of terms in Eqs. (5) and (6) (as in the $b = 1.5$ models) or completely suppress the contribution of the matter amplitudes $S_J$ via a large linear bias term $b$ and generate the observed galaxy signal $S_J^g$ with non-linear bias alone (e.g., for $b = 2$, the matter contribution to $S_4^g$ is suppressed by a factor 4). As one goes to higher orders, this procedure must be done at each order.

The predictions for the matter amplitudes $S_J$ depend slightly on the shape of the matter power spectrum $P(k)$. In Figs. 1 and 2 (solid curves) we have assumed that the galaxy spectrum measured in the APM survey is proportional to the matter spectrum. At large scales this spectral shape is similar to that of the $\Gamma = 0.2$ CDM model. For this case, Table 1 shows that $b = 1$ is compatible with no biasing, i.e. $c_2 = 0$ and $c_3 = 0$, while $b > 1$ constrains the set of possible local non-linear bias parameters.

One can also assume a different model for the matter spectrum $P(k)$, such as the standard $\Gamma = 0.5$ CDM model. In this case, it is necessary to introduce a non-local (scale-dependent) bias model, such as the cooperative galaxy formation model of Bower, etal., to explain the discrepancy in shape between the APM galaxy power spectrum and the matter density spectrum. We have shown here that scale-dependent bias alone, with no change in the overall amplitude ($b = 1$), is inconsistent with the observed higher order moments (Fig. 1 and Table 1). In principle, one could change the overall normalization of the spectrum (through a local bias) in this model; in this case, a reasonable agreement with the higher moments can be found for $b \simeq 2$ by fitting the values of $c_2$ and $c_3$. However, this choice of $b$ disagrees with the COBE DMR normalization of the standard CDM model.

Let us consider three models with initially Gaussian, scale-invariant fluctuations, to illustrate the implications of these results: standard CDM (SCDM) with $\Omega_0 = 1$, $h = 0.5$; low-density CDM (LCDM) with $\Omega_0 = 0.2$, $h = 1$; and cold plus hot dark matter (CPHDM) with $\Omega_{cold} = 0.8$, $\Omega_{hot} = 0.2$, and $h = 0.5$. The linear matter spectrum $P(k)$ in each model can be normalized using the COBE DMR measurements (Wright *et al.* 1994), which gives $\sigma_8 \simeq 1$ (SCDM, LCDM) and $\sigma_8 \simeq 0.7$ (CPHDM). The APM normalization of the galaxy spectrum corresponds roughly to $\sigma_8^g \simeq 1$, so the linear bias for each model is fixed by COBE to $b_{COBE} \simeq 1$ (SCDM, LCDM) and $b_{COBE} \simeq 1.5$ (CPHDM). The shape of the CPHDM power spectrum is qualitatively similar to that of LCDM and to that inferred for APM galaxies.

The COBE-normalized CGF model, based on SCDM and therefore with $b_{COBE} \simeq 1$, is ruled out by the $S_3$ data in the APM survey, which gives $\chi^2(4) > 11$ (Table 1). As noted above, the LCDM model, also with $b_{COBE} = 1$, is compatible with no biasing, $c_k = 0$. The COBE-normalized CPHDM model, which corresponds approximately to the $b = 1.5$, $\Gamma = 0.2$ entry in Table 1, requires a particular combination of non-linear bias parameters to fit the $S_3$ and $S_4$ data, i.e., $c_2 \simeq 0.45$, $c_3 \simeq 0.2$. This judicious choice of non-linear bias parameters would need to be extended to $c_4$ and $c_5$, using the values of $S_5$ and $S_6$ in Figure 2.

In the future, we expect that more accurate determinations of the large-scale higher order moments $S_J$ and of the multi-point functions $\xi_J(r_1, ..., r_J)$ (Fry 1994) will yield even tighter constraints on the nature of bias.

## Acknowledgements

This work was supported in part by DOE and by NASA (grant NAGW-2381) at Fermilab. E.G. acknowledges support by an HCM grant from the European Community.